\documentclass[preprint]{revtex4}
\usepackage{epsfig}

\def\beq{\begin{equation}}
\def\eeq{\end{equation}}
\def\beqa{\begin{eqnarray}}
\def\eeqa{\end{eqnarray}}

\def\GeV{\nobreak\,\mbox{GeV}}

\def\nn{\nonumber}

\newcommand{\tnk}{\Theta nK}

\newcommand{\lll}{\lambda_{\Theta}\lambda_{N}\lambda_{K}}
\def\me#1{\langle{#1}\rangle}

\def\qbar{\overline{q}}
\def\gs{g_{\rm s}}

\def\gluoncon{{\displaystyle{\gs^2G^2}}}

\newcommand{\gev}{\mbox{\rm GeV}}

\newcommand{\AmS}{{\protect\the\textfont2
  A\kern-.1667em\lower.5ex\hbox{M}\kern-.125emS}}

\begin{document}
\title{Pentaquark Decay in QCD Sum Rules}
\author{F.S. Navarra, M. Nielsen and R. Rodrigues da Silva}
\address{Instituto de F\'{\i}sica,
        Universidade de S\~{a}o Paulo, \\
        C.P. 66318,  05389-970 S\~{a}o Paulo, SP, Brazil}

\begin{abstract}
In a diquark-diquark-antiquark picture of the pentaquark we study the decay 
$\Theta \rightarrow K^{+} n$ within the framework of QCD sum rules.
After evaluation of the relevant three-point function, we extract the coupling
$g_{\Theta nK}$ which is directly related to the pentaquark width.
Restricting the decay diagrams to those with color exchange 
between the meson-like and baryon-like clusters reduces the coupling constant 
by a factor of four. 
Whereas a small decay width might be possible for a positive parity pentaquark,
it seems difficult to explain the measured width for a pentaquark with 
negative parity.

\end{abstract} 

\maketitle
\vspace{1cm}



\vspace{1cm}
\section{INTRODUCTION}

The experimental evidence for pentaquarks is now considered to be weak. However the 
strong statement that they do not exist can only be made after tests,  which are 
presently being carried on \cite{exp}.  Even if this state turns out to be unreal, its  
study generated many new theoretical ideas which can be valuable in the future  
\cite{theory}. 

One of the most puzzling characteristics of the pentaquark is its extremely
small width (much) below 10 MeV which poses a serious challenge to all theoretical
models.  Many explanations for this narrow width have been advanced \cite{theory}.  
In this work we calculate the pentaquark decay width within the framework of QCD
sum rules (QCDSR) \cite{qcdsr}. Several sum rule calculations have been performed for 
the mass of the pentaquark containing a strange quark  
\cite{zhu,mnnr,oka,eiden,morimatsu}.    
These calculations are based on two-point functions
with different interpolating currents. Surprisingly, all these determinations
give similar masses with reasonable values. 
A common problem of all determinations
is the large continuum contribution which has its origin in the high dimension
of the interpolating currents and results in a large dependence on the
continuum threshold. Another problem is the irregular behavior of the operator 
product expansion (OPE), which is  dominated by higher dimension operators and 
not by the perturbative term as it should be.

Here we present  a sum rule determination for the decay width
based on a three-point function for the decay $\Theta\to n K^+$. In this way
we can extract the coupling $g_{\Theta n K}$ which is directly related
to the pentaquark width. To describe the pentaquark we use the
diquark-diquark-antiquark model with one scalar and one pseudoscalar
diquark in a relative S-wave.

In \cite{oga,ricanari} it has been argued that such a small
decay width can only be explained if the pentaquark is a genuine 5-quark
state, i.e., it contains no color singlet meson-baryon contributions and thus 
color exchange is necessary for the decay. 
The analysis presented both in \cite{oga} and in \cite{ricanari} is only 
qualitative.   The narrowness of the pentaquark width can then be attributed to 
the non-trivial color structure of 
the pentaquark which requires the exchange of, at least, one gluon.
In this work we will also test quantitatively the hypothesis put forward in
\cite{oga,ricanari} to see whether this mechanism is sufficient 
to explain the small width.


\section{CORRELATION  FUNCTION}

The investigation of the pentaquark decay width requires a three-point function which we 
define as
\begin{eqnarray}
\label{correl}
\Gamma(p,p^{,}) &=&
\int d^4x \, d^4y  \, e^{-iqy} \, e^{ip^{,}x} \,\Gamma(x,y)\,, \nn\\
\Gamma(x,y) &=& \langle 
0|T\{\eta _N(x)j_{K}(y) \bar{\eta}_{\Theta}(0)\}|0 \rangle\,,
\end{eqnarray}
where $\eta_N$, $j_{K}$ and $\eta_{\Theta}$ are the interpolating fields associated 
with neutron, kaon and  $\Theta$, respectively \cite{ennrs}.

We next  consider the expression (\ref{correl}) in terms of hadronic degrees of freedom 
and write the phenomenological  side of the sum rule.  
Treating the kaon as a pseudoscalar particle, the interaction between the three 
hadrons is described by the following Lagrangian density:
\begin{eqnarray}
\label{lag}
{\cal L}&=&ig_{\Theta nK}\bar{\Theta}\gamma_5 Kn\quad\mbox{for $P=+$}\nn\\
{\cal L}&=&ig_{\Theta nK}\bar{\Theta}Kn\quad \ \ \ \mbox{for $P=-$}
\end{eqnarray} 
Writing the correlation function (\ref{correl}) in momentum space and inserting 
complete sets of hadronic states we obtain an expression which depends on the 
following matrix elements:
\begin{eqnarray}
\label{MatrixElements}
- i \, V(p,p') &=& < n(p',s') | \Theta(p,s) K(q) >\,, \nn\\
\langle 0|\eta _N|n(p',s')\rangle &=& \lambda_{N} u^{s'}(p')\,,\nn\\
\langle K(q)|j_{K}|0\rangle &=& \lambda_{K}\,,\nn\\
\langle \Theta(p,s)|\bar{\eta}_{\Theta}|0 \rangle &=&
\lambda_{\Theta}\bar{u}^{s}(p)\quad \ \ \ \ \ \mbox{for $P=+$}\nn\\
\langle \Theta(p,s)|\bar{\eta}_{\Theta}|0 \rangle &=&
-\lambda_{\Theta}\bar{u}^{s}(p)\gamma_{5}\quad \mbox{for $P=-$}
\end{eqnarray}
Using the simple Feynman rules derived from (\ref{lag}) we can rewrite 
$ V(p,p')$ as
\begin{eqnarray}
V(p,p') &=& - \,  g_{\Theta nK} \bar{u}^{s'}(p') \gamma_5 u^{s}(p)
\quad\mbox{ $P=+$}\nn\\
V(p,p') &=& - \, g_{\Theta nK} \bar{u}^{s'}(p') u^{s}(p)\quad\ \ \ \mbox{ $P=-$}
\end{eqnarray}
The coupling constants $\lambda_{N}$ and  $\lambda_{\Theta}$ can  be
determined from the QCD sum rules of the corresponding two-point functions.
$\lambda_{K}$ is related to the kaon decay constant through 
\begin{equation}
\label{decaek}
\lambda_{K}= \frac{f_{K}m_{K}^2}{m_u+m_s}\,.
\end{equation}
Combining the expressions above we arrive at
\begin{eqnarray}
\label{vfen3}
\Gamma_{phen} &=& 
\frac{-g_{\Theta nK}
\lambda_{\Theta}\lambda_{N}\lambda_{K}}
{(p'^2- m_{N}^2)(q^2- m_{K^{+}}^2)
(p^2- m_{\Theta}^2)} \, \nn\\
&\times& \,\Gamma _E  \,\, + \,\, \mbox{continuum}
\label{gammaphen}
\end{eqnarray}
with
\begin{eqnarray}
\Gamma _E &=&
\sigma^{\mu\nu}\gamma _{5}q_{\mu}p'_{\nu}
-im_{N}\not{\!q}\gamma _5  \nn\\
&+& i(m_N \mp m_{\Theta})\not{\!p'}\gamma _5  \nn\\
&+& i\gamma_5 (p'^2 \mp m_{\Theta}m_{N}-qp') 
\label{estrut}
\end{eqnarray}
We shall work with the 
$\sigma^{\mu\nu}\gamma _{5}q_{\mu}p'_{\nu}$ structure  because, as it was shown 
in \cite{gamma5}, 
this structure gives results which are less sensitive to the coupling scheme 
on the phenomenological side, i.e., to the choice of a pseudoscalar or 
pseudovector coupling between the kaon and the baryons.

We now come back to (\ref{correl}) and write the interpolating fields  in terms of 
quark degrees of freedom as
\begin{eqnarray}
\label{currents}
j_{K}(y)&=&\bar{s}(y)i\gamma_{5}u(y)\,,\nn\\
\eta _{N}(x)&=&\epsilon^{abc}
({d}^{T}_a(x)C\gamma _{\mu}d_{b}(x))
\gamma _5 \gamma ^{\mu}u_c(x)\,,\nn\\
\bar{\eta}_{\Theta}(0)&=&
-\epsilon^{abc}\epsilon^{def}\epsilon^{cfg}
{s}^{T}_{g}(0)C \nn\\
&\times&[\bar{d}_e(0)\gamma _5C\bar{u}^{T}_d(0)] 
[\bar{d}_b(0)C\bar{u}^{T}_a(0)]\,.
\end{eqnarray}

The pentaquark current above (proposed in \cite{oka}) contains a pseudoscalar and a scalar 
diquark. With these diquarks the two point function might receive a significant contribution 
from instantons. In \cite{dfnn} we have studied a situation in which  these instanton 
contributions affected the two-point function but gave a negligible contribution to the 
three-point function. Moreover, in \cite{cnnfd} we have observed that instantons give a
negligible contribution to heavy baryon weak decays. Motivated by these results, in this first 
calculation we shall neglect instantons.

Inserting the currents into  (\ref{correl}), the resulting expression involves 
the quark propagator in the presence of quark and gluon condensates, which is
known from previous studies. Using it  
we arrive at a final complicated expression for the correlator,  which is 
represented schematically by the sum of the diagrams of Fig. 1.

Let us consider the phenomenological side (\ref{vfen3}) and,  
following \cite{ioffe3}, rewrite it  generically as: 
\begin{equation}
\Gamma(q^2, p^2, p'^2) \, = \, \Gamma_{pp} \, + \, 
\Gamma_{pc1} \, + \, \Gamma_{pc2} \, + \, \Gamma_{cc}
\label{soma_ioffe}
\end{equation}
where  $\Gamma_{pp}(q^2, p^2, p'^2)$  stands for the pole-pole part and 
reads 
\begin{equation}
 \Gamma_{pp}  =
\frac{ - g_{\tnk}\lll}{(p^2 - m_{\Theta}^2) (p'^2 -  m^2_N) (q^2 - m^2_K)} 
\end{equation} 
The continuum-continuum term $\Gamma_{cc}$ can be obtained as usual, 
with the assumption of quark-hadron duality \cite{ennrs}.

The pole-continuum transition terms are contained in $\Gamma_{pc1}$ and
$\Gamma_{pc2}$. They can be explicitly written as a double dispersion integral:
\begin{eqnarray}
\label{gammapc1}
\Gamma_{pc1} &=& \int_{m^2_{K^{*}}}^{\infty}  
\, \frac{b_1(u,p^2)  \,  du}
{(m^2_N  - p'^2) (u - q^2)}\,,\nn\\
\Gamma_{pc2} &=&  \int_{m^2_{N^{*}}}^{\infty} \,
 \frac{b_2(s,p^2) \,  d s}
{(m^2_K - q^2) (s -  p'^2)}\,.
\end{eqnarray}

Since there is no theoretical tool to calculate the unknown
functions $b_1(u,p^2)$ and $b_2(s,p^2)$ explicitly, one has to employ a
parametrization for these terms. We will use two different parametrizations:
one with a continuous function for the $\Theta$ and one where the pole term
is singled out.

We shall assume that the functions $b_1$ and $b_2$ have the following form: 
\begin{eqnarray}
b_1(u,p^2) &=& \widetilde{b_1}(u) \int^{\infty}_{m_{\Theta}^2} d \omega
\frac{b_1(\omega)}{\omega - p^2}    \nonumber \\
b_2(s,p^2) &=& \widetilde{b_2}(s) \int^{\infty}_{m_{\Theta}^2} d \omega
\frac{b_2(\omega)}{\omega - p^2} 
\label{b1b2}
\end{eqnarray}
with continuous functions $b_{1,2}(w)$, starting from $m^2_\Theta$. This is our
{\bf parametrization A}.
The functions $\widetilde{b_1}(u)$ and $\widetilde{b_2}(s)$ describe the excitation 
spectra of the kaon and the nucleon, respectively.  After Borel transform,
the pole-continuum term contains one unknown constant factor which can be determined
from the sum rules.

In order to investigate the role played by the 
$\Theta$ continuum, we shall now explicitly force the 
phenomenological side to contain only the pole part of the $\Theta$, both in the pole-pole 
term and in the pole-continuum terms. 
This can formally  be done by choosing
$b_1(\omega) = b_2(\omega) = \delta(\omega - m_{\Theta}^2)$ in (\ref{b1b2}) and
the functions then read:
\begin{eqnarray}
b_1(u,p^2) &=& \frac{\widetilde{b_1}(u)}{m_\Theta^2-p^2}\,,\nn\\
b_2(s,p^2) &=& \frac{\widetilde{b_2}(s)}{m_\Theta^2-p^2}\,.
\end{eqnarray}
This is our {\bf parametrization B}.
In this case we have the $\Theta$ in the ground state. Again, in the final expressions 
this gives  additional constants which can
be calculated.


\section{SUM  RULES}

The sum rule  may be written identifying the phenomenological and theoretical descriptions 
of the correlation function. As mentioned above, 
we shall work with the $\sigma^{\mu\nu}\gamma _{5}q_{\mu}p'_{\nu}$ structure. 
In  the case of the three-point function considered here, there are two independent momenta 
and we may perform either a single or a double Borel transform.    
We first consider the choice:
\begin{equation}
\mbox{(I)} \,\,\,\,\,\,\,\,\,\,   q^2 \, = \, 0  \,\,\,\,\,\,\,\,\   p^2 \, = \, p'^2   
\end{equation}
and perform a single Borel transform: $p^2 \, = \, - P^2$ and $ P^2 \rightarrow M^2$. In this
case we take $m_K^2  \simeq 0$ and single out the $1/q^2$-terms. The second choice is:
\begin{equation}
\mbox{(II)} \,\,\,\,\,\,\,\,\,\, q^2 \, \neq  \, 0  \,\,\,\,\,\,\,\,\   p^2 \, = \, p'^2\,. 
\end{equation}
Here we perform two Borel transforms: $p^2 \, = \, - P^2$ and $ P^2 \rightarrow M^2$ 
and also $q^2 \, = \, - Q^2$ and $ Q^2 \rightarrow M^{'2}$. 
We have also considered the choice $ q^2 \, = \, p^2 \, = \, p'^2 \, = - P^2 $, 
performing one single Borel transform ($ P^2 \rightarrow M^2$).  
However, in the present calculation 
we were not able to find  a stable sum rule. Introducing the notation  
$G = - g_{\Theta nK}\lambda_{\Theta}\lambda_{N}\lambda_{K}$  
and  using (I) and (II) we obtain the following sum rules:

{\bf Method I:}
\begin{equation}
\label{soma1}
\Gamma_{pp}(M^2)+\Gamma_{pc2}(M^2)=
\int_0^{s_0}  ds  \, \rho_{th}(s) \, e^{-s/M^{2}}
\end{equation}
with
\begin{equation}
\Gamma_{pp}(M^2) = G \frac{e^{-m^2_{\Theta}/M^{2}}  
-e^{-m^2_{N}/M^{2}}  
}{m^2_{\Theta} - m^2_N} 
\end{equation}
and  for the pole-continuum part we obtain
\begin{eqnarray}
\Gamma_{pc2}(M^2) &=& A \,  e^{-m^2_{N^{*}}/M^{2}}
\quad \mbox{param. A}\nn\\
\Gamma_{pc2}(M^2) &=& A \,  e^{-m^2_\Theta/M^{2}}
\quad \ \mbox{param. B}
\end{eqnarray}
In both parametrizations the term $\Gamma_{pc1}$ is exponentially suppressed
and, as discussed in \cite{ioffe3}, has been neglected. $A$ is an unknown 
constant and can be determined from the sum rules.

{\bf Method II}
\begin{equation}
\label{soma2}
\Gamma_{pp}(M^2,M^{'2})+\Gamma_{pc2}(M^2,M^{'2}) =
\end{equation}
\begin{equation}
\int_0^{u_0} \, du \, \int_0^{s_0} \, ds \,  \rho_{th}(s,u) 
\, e^{-s/M^{2}} \, e^{-u/M^{'2}}
\end{equation}
with
\begin{equation}
\Gamma_{pp} = G \,  e^{-m^2_K/M^{'2}} \,
\frac{e^{-m^2_{\Theta}/M^{2}}  
-e^{-m^2_{N}/M^{2}}}  
{m^2_{\Theta} - m^2_N} 
\end{equation}
and with
\begin{equation}
\Gamma_{pc2} = A \, e^{-m^2_K/M^{'2}} \, 
e^{-m^{2}_{N^*}/M^{2}}
\end{equation}
for parametrization A and 
\begin{equation} 
\Gamma_{pc2} = A \, e^{-m^2_K/M^{'2}} \, 
e^{-m^{2}_{\Theta}/M^{2}}
\end{equation}
for parametrization B.  
Also in this case $\Gamma_{pc1}$ is exponentially suppressed. In the 
above expressions $\rho_{th}$ is the double discontinuity computed directly 
from the theoretical (OPE) description of the correlation function (see 
\cite{ennrs} for details and also \cite{bclnn}) and $s_0$ is the continuum 
threshold of the nucleon defined as $s_0  \, = \, ( m_N \, + \, \Delta_N )^2 $.


\section{RESULTS} 

The hadronic masses are
$m_N = 938$ MeV, $m_{N^*}=1440$ MeV,  $m_K = 493$ MeV and $m_{\Theta} = 1540$ MeV.  
For each of the sum rules above (Eqs. (\ref{soma1}) and (\ref{soma2}))   
we can take the derivative with respect to $1/M^2$ and in this way 
obtain a second sum rule. In each case 
we have thus a system of two equations and two unknowns ($G$ and $A$) which 
can then be easily solved. 

In the numerical analysis of the sum rules we use the following values for the 
condensates:  $\me{\qbar q}=\,-(0.23\pm 0.02)^3\,\GeV^3$,
$\langle\overline{s}s\rangle\,=0.8 \, \me{\qbar q}$, $ <\bar{s} g_s {\bf\sigma.G}s > 
= m_0^2 \, \me{\bar{s}s}$ with $m_0^2=0.8\,\GeV^2$ and $\me{\gluoncon}=0.5~\GeV^4$.
The gluon condensate has a large error of about a factor 2, but its influence
on the analysis is relatively small. 
The couplings constants $\lambda _{N}$ and $\lambda _{\Theta}$ are taken 
from the corresponding two-point functions:
\begin{equation}
\lambda _{N}= (2.4 \pm 0.2)  \times 10^{-2} \, \mbox{GeV}^3 
\end{equation}
\begin{equation}
\lambda _{\Theta}= (2.4 \pm 0.3) \times 10^{-5} \, \mbox{GeV}^6
\label{lambdas}
\end{equation}
The coupling $\lambda _{K}$ is obtained from (\ref{decaek}) with $f_K = 160$ MeV,
$m_s = 100$ MeV and $m_u = 5$ MeV: 
\begin{equation}
\lambda _{K}=0.37 \, \mbox{GeV}^2 \,.
\label{lambdak}
\end{equation}

In Fig. 1, among these OPE diagrams there are two distinct subsets. 
In the first 
(from 1a to 1g) there is no gluon line connecting the petals and therefore 
no color 
exchange. A diagram of this type we call color-disconnected. In the second 
subset 
of diagrams (1h, 1i and 1j) we have color exchange. If there is no color 
exchange, the 
final state containing two color singlets was already present in the 
initial state, 
before the decay, as noticed in \cite{morimatsu}. 
In this case the pentaquark had a component similar to a  $K-n$ molecule. 
In the second case the pentaquark was a genuine 5-quark  state with a
non-trivial color structure.
We may call this type of diagram a color-connected (CC) one. In our analysis we 
write sum rules for both cases: all diagrams and only color-connected. The former 
case is standard in QCDSR calculations and therefore we shall omit details and 
present only the results. The latter case implies that 
 the pentaquark is a genuine
5-quark state and the evaluation of $g_{\Theta nK}$ will thus be based
only on the CC diagrams. We shall work in the Borel window given by 
$1\,\gev^2 < M^2,M'^2 < 1.5 \,\gev^2$.
Since the strange mass is small, the dominating diagram is Fig. 1b
of dimension three with one quark condensate. 
In the range considered, the dimension 5 condensates are substantially
suppressed compared to this term. 

We have found out that the contribution from the pole-continuum part is
of a similar size as the pole part. For lower values of $M^2$ around
1 GeV$^2$, the pole contribution dominates, however, for larger values
of $M^2$ the importance of the pole-continuum contribution grows and
eventually becomes larger than the pole part. This is an additional
reason to restrict the analysis to small values for the Borel parameters.

We have evaluated  the  sum rules for the coupling constant 
computed with all diagrams of Fig. 1 and we have found that they are very stable.   
We give  the values of the coupling 
extracted at $M^2 = 1.5$ GeV$^2$ and  $M'^2 = 1$ GeV$^2$  in Table I.  
In what follows we shall present our results for the coupling constant 
$g_{\Theta n K}$ obtained with the color connected diagrams only.  
In Fig. 2 we show the coupling, given 
by the solution of the sum rule  I A (\ref{soma1}), as a function of the 
Borel mass squared $M^2$. 
Different lines show different values of the continuum threshold $\Delta_N$. As 
it can be seen, $g_{\Theta n K}$ is remarkably stable with respect to variations both in
$M^2$ and in $\Delta_N$. In Fig. 3 we show the coupling obtained with the sum
rule  II A  
(\ref{soma2}). We find again fairly stable results which are very weakly
dependent on the continuum threshold.  In Fig. 4 we show the results of the 
sum rule  I B. 
In Fig. 5 we present  the result of the sum rule  II B. The meaning of  the
different lines is the same as in the previous figures. 
The results are similar to the cases before.

In Table I we present a summary of our results for $g_{\Theta n K}$ giving
emphasis to the difference between the results obtained with all diagrams and with 
only the color-connected ones. For the continuum thresholds we have 
employed $\Delta_N = \Delta_K = 0.5$ GeV. 
\vspace{0.5cm}

\begin{center}
\small{Table I:$g_{\Theta nK}$ for various cases}\\
\end{center}
\begin{center}
\begin{tabular}{|c|c|c|}  \hline
case & $|g_{\Theta nK}|$ (CC) &  $|g_{\Theta nK}|$ (all diagrams)  \\
\hline
\hline
I A & 0.71 & 2.59  \\
\hline
II A & 0.82 & 3.59   \\
\hline
I B & 0.84 & 3.24  \\
\hline
II B & 0.96  & 4.48  \\
\hline
\end{tabular}
\end{center}

\vspace{0.5cm}

For our final value of $g_{\Theta nK}$ we take an average of the
sum rules  I A - II B. It is interesting to observe that the influence of
the continuum threshold is relatively small, especially when compared to
the corresponding two-point functions.

Considering the uncertainties in the continuum thresholds, in the
coupling constants $\lambda_{K,N,\Theta}$ and in the quark condensate we  
get an uncertainty of about 50\%. Our final result then reads:
\begin{eqnarray}
  |g_{\Theta nK}| \mbox{(all diagrams)} &=& 3.48 \, \pm 1.8\,,\nn\\
  |g_{\Theta nK}| \mbox{(CC)}&=& 0.83 \, \pm 0.42\,.
\label{gfinal}
\end{eqnarray}
Including all diagrams, the prediction for $\Gamma_\Theta$ is
then 13 MeV (652 MeV) for a positive (negative) parity pentaquark.
In the CC case we get a width of 0.75 MeV (37 MeV) for a positive (negative) 
parity pentaquark.  The measured upper limit of the width is around 5-10 MeV 
both in the Kn channel (considered here) and in the Kp channel.

We see that it is very difficult to obtain the measured decay width for a negative
parity pentaquark.


\section{SUMMARY AND CONCLUSIONS}

We have presented a QCD sum rule study of the decay of the 
$\Theta^{+}$ pentaquark  using a  diquark-diquark-antiquark scheme
with one scalar and one pseudoscalar diquark. 
Based on the  evaluation of the relevant three-point function, 
we have  computed the coupling  constant $g_{\Theta nK}$. 
In the operator product expansion we have included all diagrams
up to dimension 5. In this particular type of sum rule a complication
arises from the pole-continuum transitions which are not exponentially
suppressed after Borel transformation and must be explicitly included.
The analysis was made for two different pole-continuum parametrizations
and in two different evaluation schemes. The results are consistent with
each other. In addition, we have tested the ideas presented in
\cite{oga,ricanari} by including only diagrams with color exchange.
Our final results are given in eq. (\ref{gfinal}).

We find that for a positive parity pentaquark a width 
much smaller than 10 MeV would indicate a pentaquark which contains no 
color-singlet meson-baryon contribution. For a negative parity pentaquark, 
even under the assumption that it is a genuine 5-quark state, we can
not explain the observed narrow width of the $\Theta$.

\smallskip

{\bf Acknowledgements:} It is a pleasure to thank R.D. Matheus, M. Karliner and
A. Zhitnitsky for instructive discussions. This work has been supported by CNPq 
and  FAPESP (Brazil). 
\vspace{0.5cm}





\begin{figure} \label{fig2}
\centerline{\psfig{figure=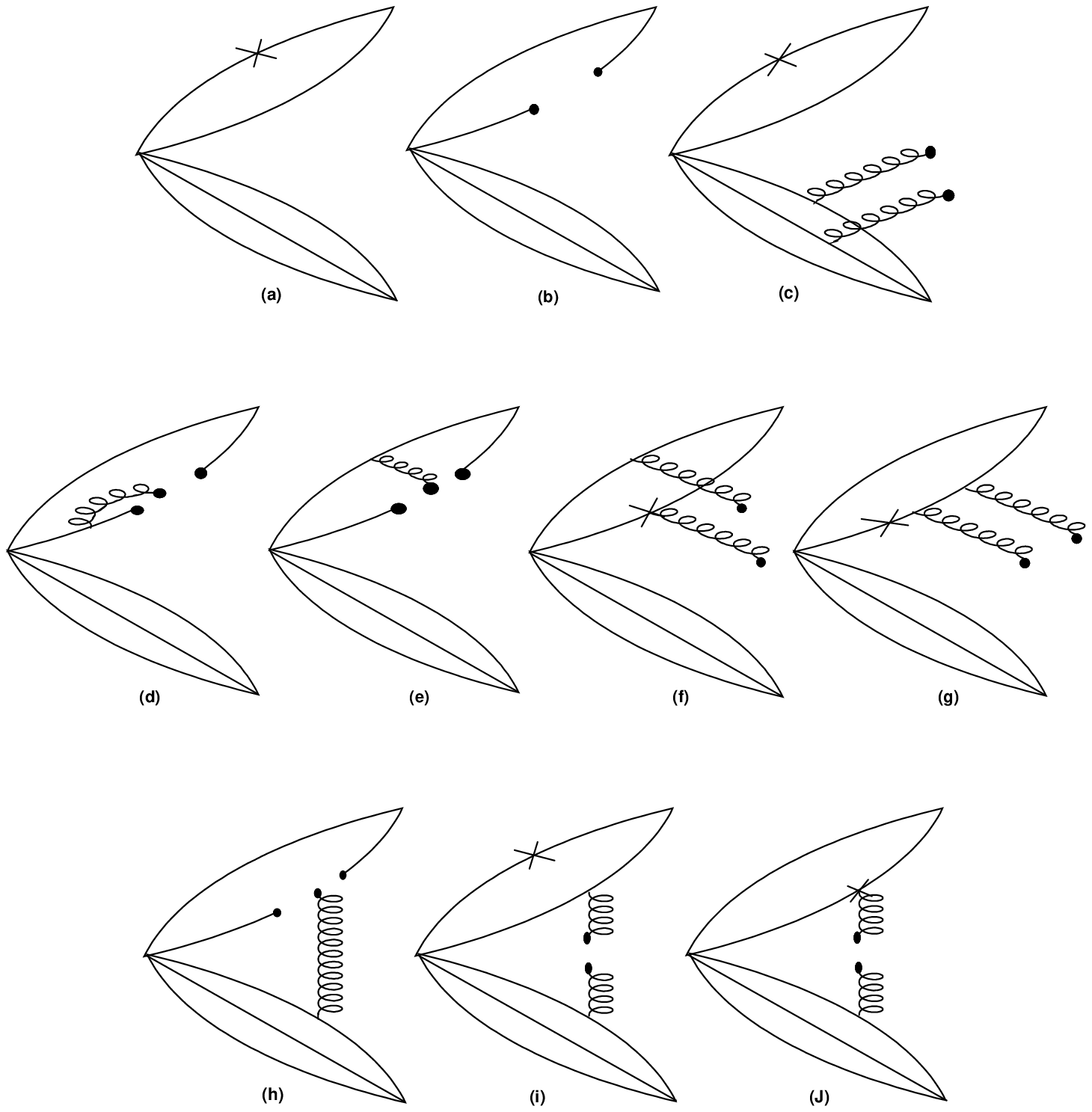,width=8cm,angle=0}}
\caption{The main diagrams which contribute to the theoretical side 
of the sum rule in the relevant structure.
a) - g) are the color-disconnected diagrams, whereas h) - j) are the 
color-connected diagrams. The cross indicates the insertion of
the strange mass.}
\end{figure}

\begin{figure} \label{fig3}
\centerline{\psfig{figure=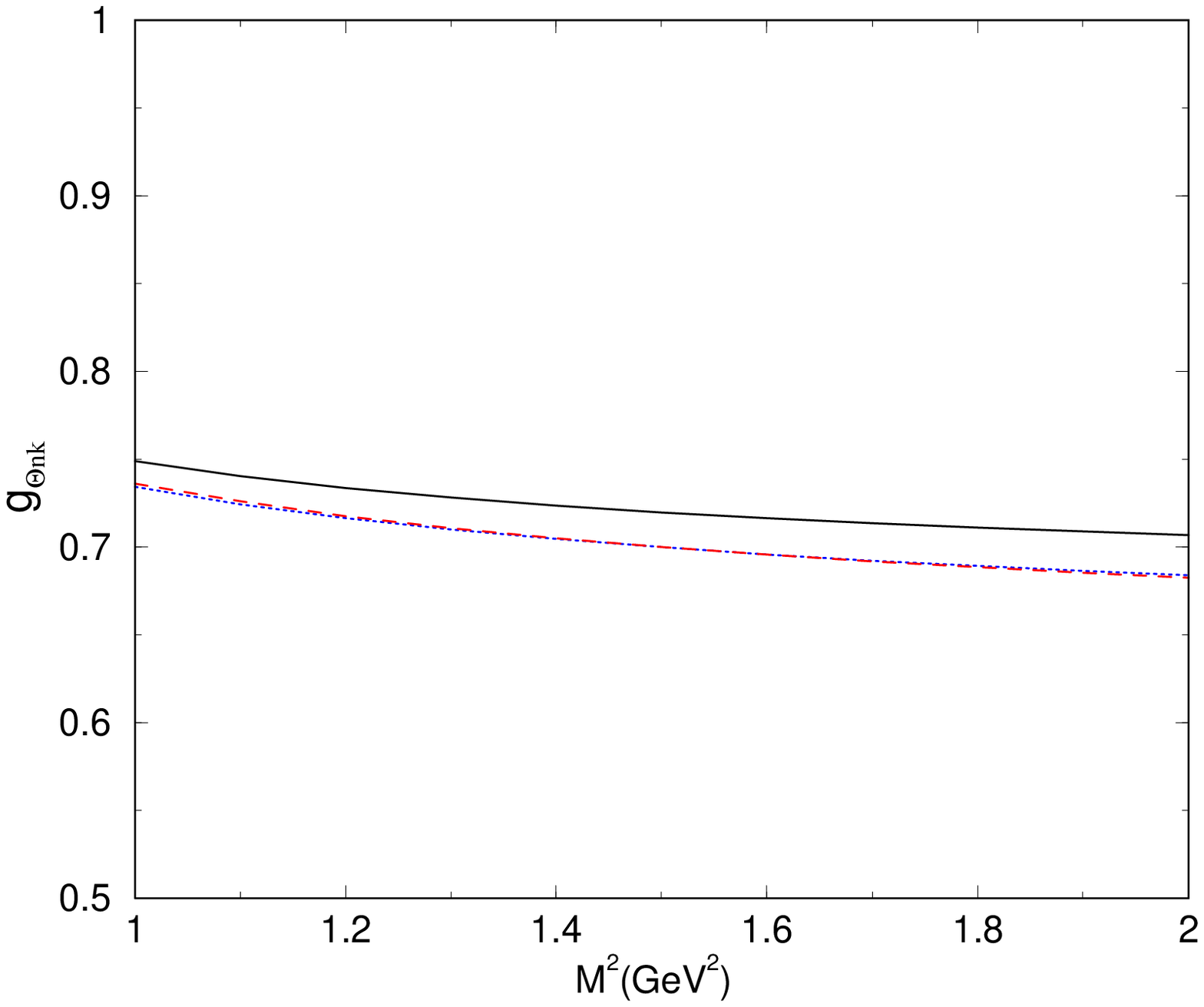,width=7cm,angle=0}}
\caption{$|g_{\tnk}|$ in case  I A with three different continuum threshold parameters. 
Solid line: $\Delta_N=0.5$ GeV, dotted line:$\Delta_N=0.4$ GeV, 
dash-dotted line:  $\Delta_N=0.6$ GeV.}
\end{figure}

\begin{figure} \label{fig4}
\centerline{\psfig{figure=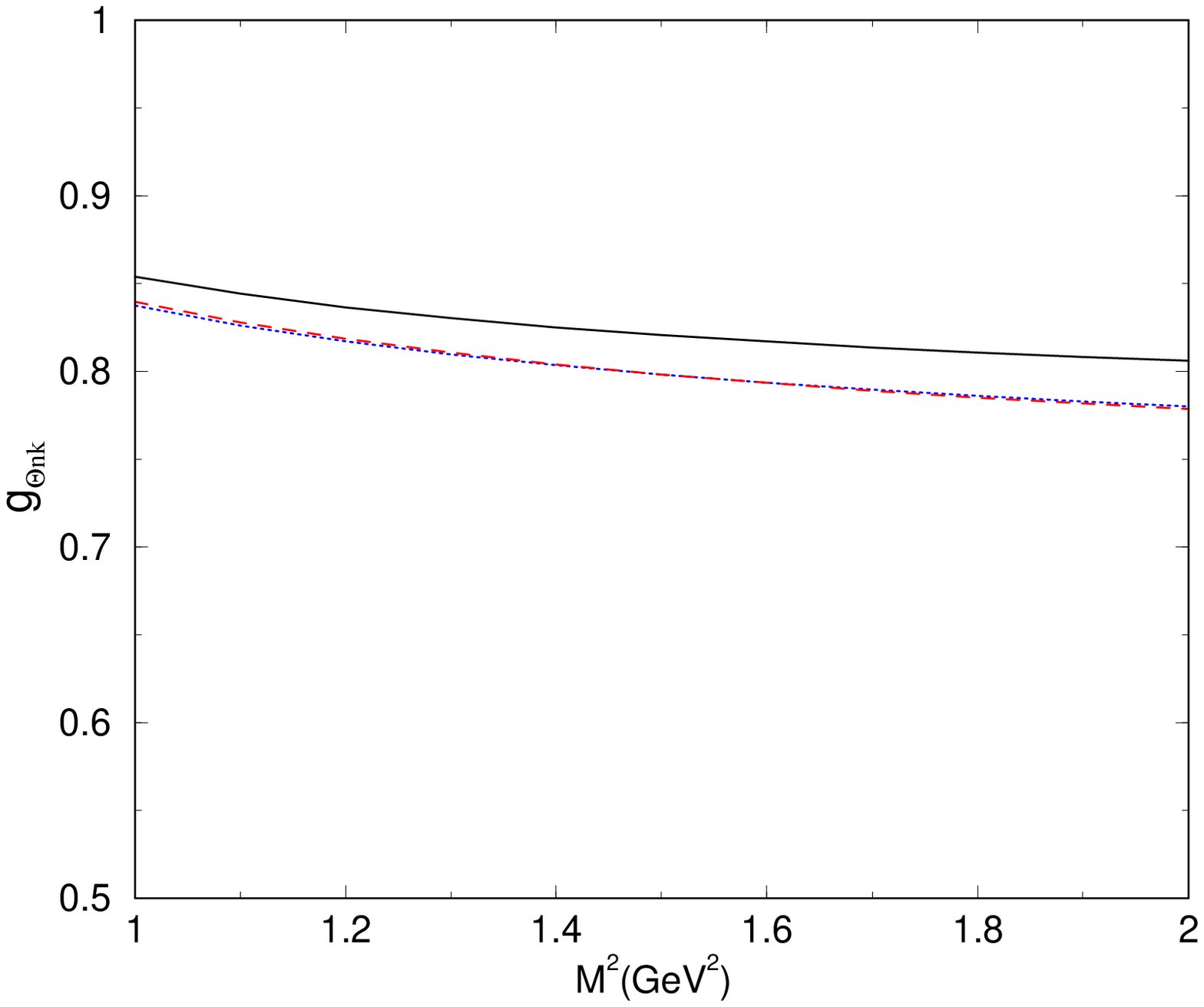,width=7cm,angle=0}}
\caption{$|g_{\tnk}|$ in case  II A. Solid line:  $\Delta_N=0.5$ GeV. 
Dotted line: $\Delta_N=0.4$ GeV. Dashed line: $\Delta_N=0.6$ GeV.
$M^{'2}=1$ GeV$^{2}$.} 
\end{figure}

\begin{figure} \label{fig5}
\centerline{\psfig{figure=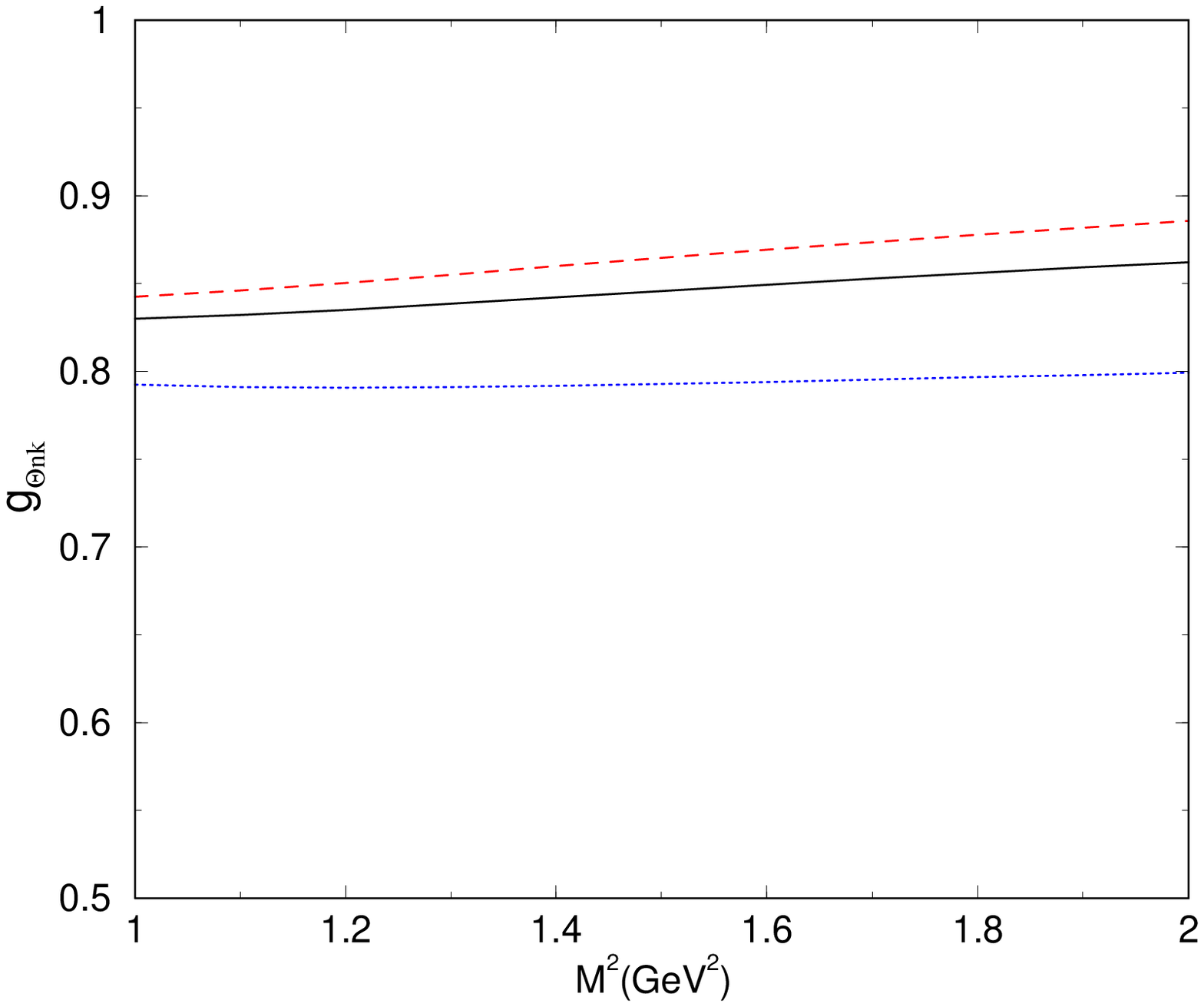,width=7cm,angle=0}}
\caption{$|g_{\tnk}|$ in case  I B  with three different continuum threshold parameters. 
Solid line: $\Delta_N=0.5$ GeV, dotted line:$\Delta_N=0.4$ GeV, 
dash-dotted line:  $\Delta_N=0.6$ GeV.}
\end{figure}

\begin{figure} \label{fig6}
\centerline{\psfig{figure=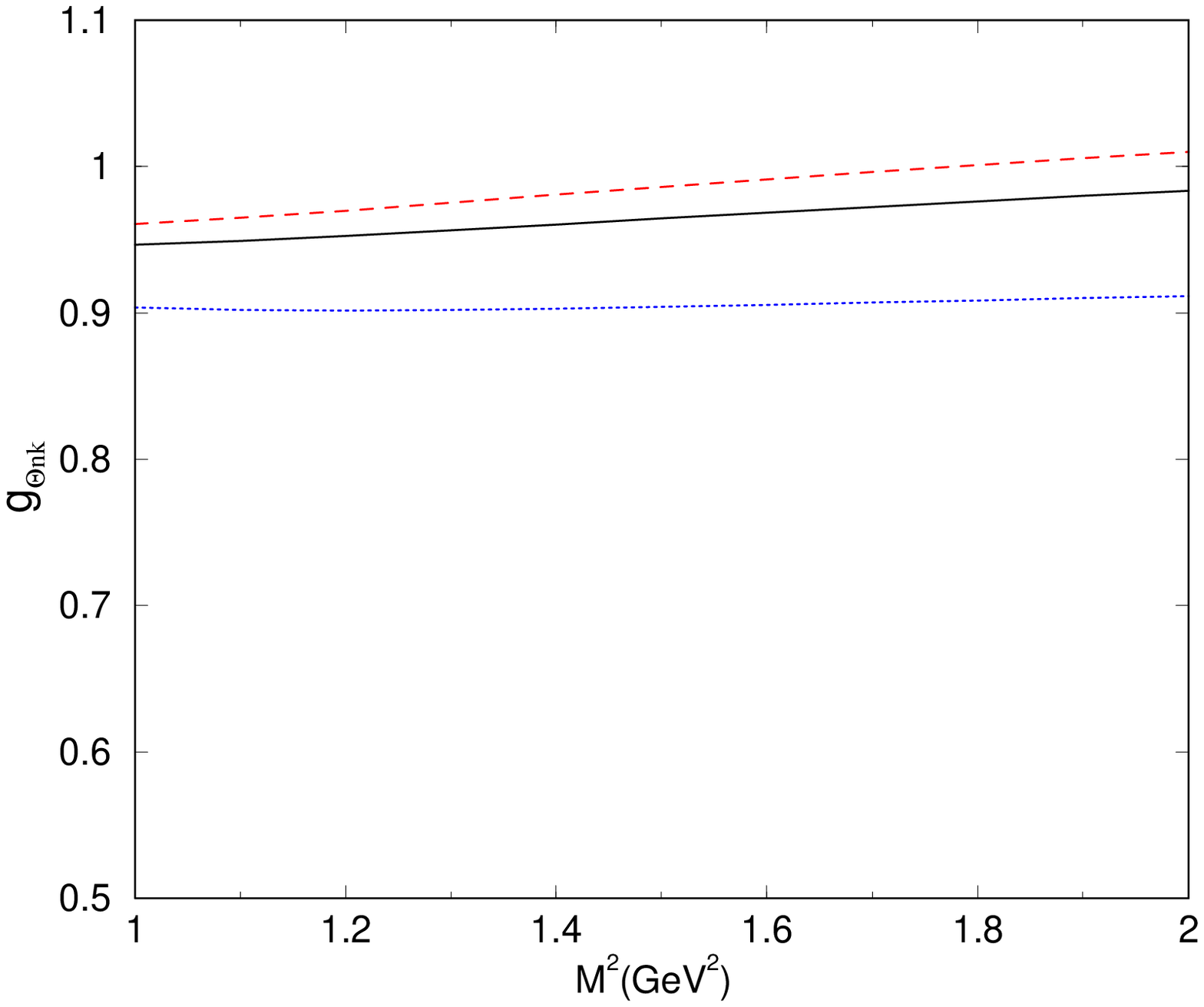,width=7cm,angle=0}}
\caption{$|g_{\tnk}|$ in case  II B. Solid line:  $\Delta_N=0.5$ GeV. 
Dotted line: $\Delta_N=0.4$ GeV. Dashed line: $\Delta_N=0.6$ GeV.
$M^{'2}=1$ GeV$^{2}$.} 
\end{figure}

\end{document}